# Two-Dimensional Bipolar Junction Transistors

**Behnaz Gharekhanlou, Sina Khorasani, and Reza Sarvari**

School of Electrical Engineering, Sharif University of Technology, P. O. Box 11365-9363, Tehran, Iran
Email: khorasani@sina.sharif.edu; fax: +98-21-6602-3261

**Abstract**
Recent development in fabrication technology of planar two-dimensional (2D) materials has brought up possibilities of numerous novel applications. Our recent analysis has revealed that by definition of p-n junctions through appropriate patterned doping of 2D semiconductors, ideal exponential I-V characteristics may be expected. However, the theory of 2D junctions turns out to be very much different to that of the standard bulk junctions. Based on this theory of 2D diodes, here we construct for the first time, a model to describe the 2D Bipolar Junction Transistors (2D-BJTs). We derive the small-signal equivalent model, and estimate the performance of a 2D-BJT device based on Graphone as the example material. A current gain of about 138 and maximum threshold frequency of 77GHz, together with a power-delay product of only 4fJ per 1μm lateral width is expected at an operating voltage of 5V. Also, we derive necessary formulae and a new approximate solution for continuity equation in the 2D configuration, which have been verified against numerical solutions.

## 1. Introduction

Since the discovery of Graphene in 2004 [1], lots of attention has been drawn to the so-called two-dimensional (2D) materials. The common particular aspect of 2D materials is that they are all only one monatomic layer thick.

In general, basic properties of 2D materials can be routinely engineered by doping, functionalisation, and chemical modification [1-6], such as hydrogenation or oxidization. For instance, the fully-hydrogenated Graphene, or Graphane, should be obtainable as a stable 2D hydrocarbon [7,8]. Since then, a great deal of research has been focused on the properties of this unique material, investigating its electronic, optical [9], nanotube allotropes [10], and even superconductive properties [11]. It has been shown that although perfectly-ordered Graphane may be difficult to obtain [12], however, as it has been shown [13], the influence of substrate material on the quality of 2D crystal domains cannot be neglected.

Recently, a similar material with identical chemical configuration has been created by hydrogenation of Germanene, named as Germanane [14]. These monolayer hydrogenated dielectrics are very much similar and are expected to be described by identical theories. For most applications such as ours, single-sided hydrogenation of Graphene is however much better suited, commonly referred to as Graphone [13].

It has been shown that stacks of 2D materials could also lead to unprecedented applications in everyday's science and technology. With the advent of insulating 2D Boron Nitride, further properties of such layered 2D sandwichs have been proposed [13,15].

As we have shown using tight binding calculations Graphane has potential advantages over Graphene with regard to having a controllable bandgap which makes it highly desirable for integrated electronic applications [1,16,17]. Many articles are concerned with simulations of material properties only, whereas studying the operation of the proposed class of 2D structures would need a significantly different theory.

The formulation discussed in this article is based on the theory of 2D p-n rectifying junction diodes developed in an earlier study of the authors [18]. In the previous research, we investigated a p-n junction with abrupt doping profile made of a Graphone sheet [18]. Evidently, a pair of similar 2D junctions may provide a fairly good basis for operation of a 2D Bipolar Junction Transistors (2D-BJTs). We actually had made a basic study of 2D-BJTs [19], however, our initial effort led to a device with a current gain less than unity.

Here, we revisit the design of a 2D-BJT with abrupt doping profile, and Graphone is used as the host 2D material. An electrostatic analysis of the transistor is done together with Shockley's law. The small-signal equivalent circuit of the 2D-BJT in its active region is obtained, and we show that extremely large output resistances together with vanishingly small power-delay products are possible at once, thus bringing the ultralow power circuits one step closer to the reality.

For the first time, we derive the necessary formulae for description of 2D-BJTs, study the 2D continuity equations, and suggest useful approximate analytical solutions. Accuracy of approximate analytical solutions is established by comparing to the exact numerical solutions. We anticipate that 2D-BJTs based on Graphane, Germanane, and similar 2D materials, would also share similar theories.

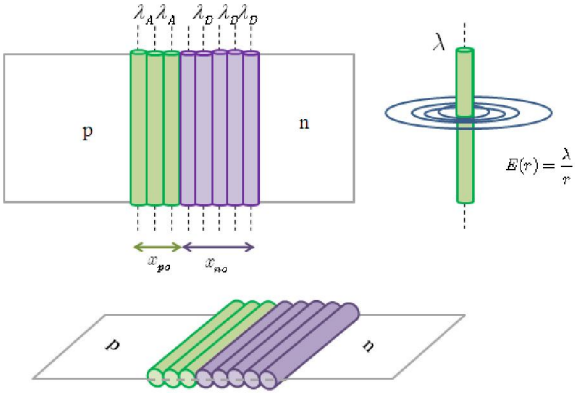

**Figure 1.** New charge model to illustrate surface charge.

## 2. Physical Parameters

Some of the required physical parameters are already reported in literature, and the remaining data are derived from band structure and/or available relations. These include the effect of dopant type and concentration of carrier mobility, effective dielectric constant, effective mass and the 2D density of states.

There exist experiments which have probed the influence of chemical dopants on the carrier mobility in Graphene [20]. Hall measurements revealed that $NO_2$ and $H_2O$ adsorption act as acceptors whereas $NH_3$ and $CO$ are donors. It is reported that when Graphene is treated with atomic hydrogen, it starts exhibiting an insulating behavior because of increasing band gap. This is while the electron mobility of Graphane [8] has been also measured in experiment, and it is found that by annealing, the mobility of Graphane is recovered to the Graphene mobility 3,500cm$^2$/Vs. Also, recent experiments have reported the influence of chemical dopants on the carrier mobility in Graphene [20,21]. It is observed that even for surface chemical dopant concentrations in excess of $10^{12}$cm$^{-2}$ there is no observable change in the carrier mobility.

We have found the effective mass and 2D density of states from Graphone band structure [18]. For this purpose, we used the tight-binding method to obtain band structure, leading to the effective masses $m_C^* = 1.03$, $m_V^* = 0.63$ [18]. Based on these data, the total planar density of states was obtained to be $N_{C(V)}^{2D} = m_{C(V)} k_B T / \pi \hbar^2$ [18].

The relative dielectric constant of Graphene is also reported to be about 2.5 [22]. There is however no available measurement on the dielectric constant of Graphane. In this study, we use the dielectric image method and take on a value of 2.7 for the relative dielectric constant of Graphane layer on $SiO_2$ substrate. It is not difficult to show using the electrostatic image method that the effective permittivity seen by charges lying on an interface between two dielectrics is simply $\varepsilon_{eff} = (\varepsilon_2 + \varepsilon_1)/2\varepsilon_1$. Hence, if the substrate is $SiO_2$ with the relative permittivity 4.4 and the covering layer on the Graphane is vacuum or air, then the effective permittivity of the layer will be given by $\varepsilon_{eff} = (4.4 + 1)/2 = 2.7$.

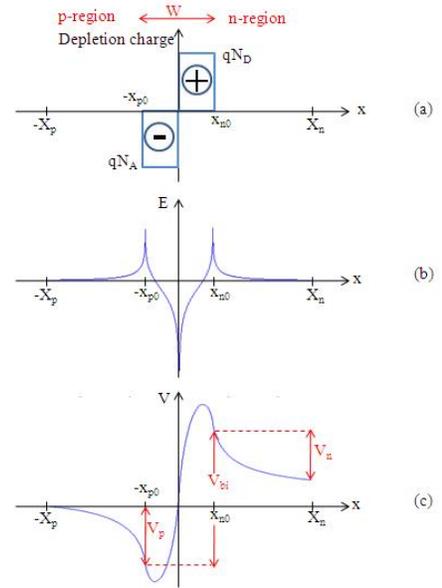

**Figure 2.** Abrupt p-n junction in the thermal equilibrium: (a) Space charge distribution; (b) Electric field distribution; (c) Potential distribution.

## 3. The 2D P-N Junction

In our previous work we designed an abrupt p-n junction with arbitrary dimensions. We assumed that the junction has infinity length and solved the continuity equation with the continuity conditions at the two edges. Now we require designing an abrupt p-n junction with submicron dimensions. This requirement will be needed to obtain high gain and bandwidth. For this purpose, we have to resolve the continuity equation for a limited length junction. Also we will drive the necessary theoretical formulae and compare them with numerical calculations.

### 3.1. Built-in Potential and Depletion Layer

In a bulk 3D p-n junction we deal with space charge within the depletion region. But for a 2D junction, there is a surface charge. So we have introduced a new charge model to illustrate this 2D system [18]. We divide this charge sheet into infinitely many thin line charges with differential line charge densities $\lambda_A$ and $\lambda_D$ in p and n depletion regions, respectively. Figure 1 shows this idea schematically. The electric field due to each line charge is $E(r) = \lambda/r$ where *r* is measured with respect to the symmetry axis of the line charge. The total electric field had been reported for the four regions over the surface of the junction in [18], which may be now combined into one unified equation as

$$E(x) = \mathbf{E}(x, z=0) \cdot \hat{x} = \lambda_A \ln\left|\frac{x}{x+x_{p0}}\right| - \lambda_D \ln\left|\frac{x_{n0}-x}{x}\right| \quad (1)$$

where $\lambda_{A,D} = qN_{A,D}/2\pi\varepsilon$. Figure 2 shows this abrupt p-n junction under the thermal equilibrium. It may be noticed that since $\partial \mathbf{E}(x, z=0) \cdot \hat{z}/\partial z \neq 0$, the derivative $\partial E(x)/\partial x$ does not reproduce the initial charge distribution, unlike the regular step junctions. This is simply because $\partial E(x)/\partial x \neq \nabla \cdot \mathbf{E}(x, z=0)$. Now, from the Poisson equation, the potential distribution is obtained. The potential across the depletion region, gives the built-in potential

$$V_{bi} = (\lambda_A + \lambda_D)\left[x_{p0}\ln(W/x_{p0}) + x_{n0}\ln(W/x_{n0})\right] \quad (2)$$

where $W = x_{n0} + x_{p0}$. For a symmetric abrupt junction, $x_{n0} = x_{p0} = W/2$, $\lambda_A = \lambda_D = \lambda$. Hence (3) is reduced to

$$V_{bi} = \lambda W \ln 2 \quad (3)$$

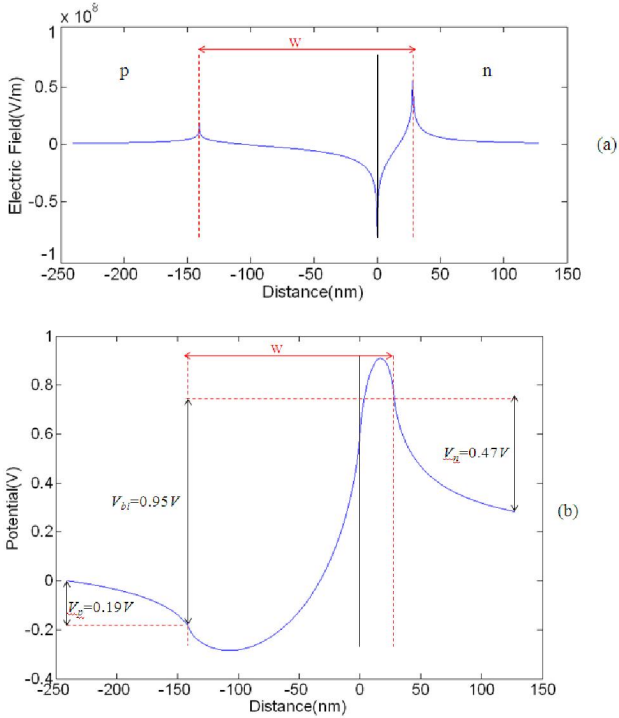

**Figure 3.** (a) Electric field distribution; (b) Electric potential distribution for the designed p-n junction.

It can be seen that the built in potential is proportional to $W$ while for a conventional bulk p-n junction it is proportional to $W^2$. Now since the electric field is non-zero inside the p and n neutral regions, the potential drops across these regions are found as

$$\begin{aligned}V_p = &-(\lambda_A + \lambda_D)\left[x_{p0}\ln(x_{p0}) + X_p\ln(X_p)\right]\\ &+ \lambda_A(X_p - x_{p0})\ln(X_p - x_{p0}) - \lambda_D W \ln(W)\\ &+ \lambda_D(X_p + x_{n0})\ln(X_p + x_{n0}), \quad -X_p < x < -x_{p0}\end{aligned} \quad (4)$$

$$\begin{aligned}V_n = &(\lambda_A + \lambda_D)\left[x_{n0}\ln(x_{n0}) - X_n\ln(X_n)\right]\\ &+ \lambda_A(X_n + x_{p0})\ln(X_n + x_{p0}) - \lambda_A W \ln(W)\\ &+ \lambda_D(X_n - x_{n0})\ln(X_n - x_{n0}), \quad x_{n0} < x < X_n\end{aligned} \quad (5)$$

To simplify the analysis, the depletion approximation is here used. Hence, under the thermal equilibrium, the total negative charge per unit width in the p-side must be equal to the total negative charge per unit width in the n-side

$$N_A x_{p0} = N_D x_{n0} \quad (6)$$

For an abrupt junction the built-in potential is equal to

$$V_{bi} \approx V_T \ln\left(N_A N_D/n_i^2\right), \quad V_T = kT/q \quad (7)$$

From (2), (6) and (7), the depletion width is calculated to be

$$W = \frac{2\pi\varepsilon}{q} \times \frac{V_T \ln\left(N_A N_D/n_i^2\right)}{N_D \ln\left(1 + N_A/N_D\right) + N_A \ln\left(1 + N_D/N_A\right)} \quad (8)$$

### 3.2. Current-Voltage Characteristics

The injected minority-carriers distribution in the n-side is governed by the continuity conditions at the two edges, subject to the Shockley's boundary conditions

$$\begin{aligned}&-\frac{p_n(x) - p_{n0}}{\tau_p} + \mu_p(-E(x)\frac{d}{dx} + V_T\frac{d^2}{dx^2})p_n(x) = 0\\ &p_n(x = x_{n0}) = p_{n0}\exp(V/V_T)\\ &p_n(x = X_n) = p_{n0}\end{aligned} \quad (9)$$

and similarly in the p-side

$$\begin{aligned}&-\frac{n_p(x) - n_{p0}}{\tau_n} + \mu_n(E(x)\frac{d}{dx} + V_T\frac{d^2}{dx^2})n_p(x) = 0\\ &n_p(x = -x_{p0}) = n_{p0}\exp(V/V_T)\\ &n_p(x = -X_p) = n_{p0}\end{aligned} \quad (10)$$

As mentioned before, in the neutral region the electric field is non-zero and $E(x)$ represents the local electric field of each point in this region.

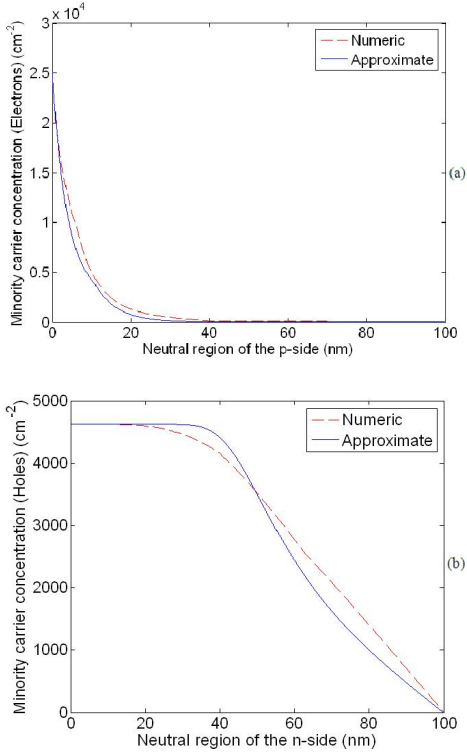

**Figure 4.** Minority carrier distributions under forward bias: (a) Hole concentration in the n-region; (b) Electron concentration in the p-region.

To solve these equations numerically, besides using these equations directly, we can also use the logarithmic form discussed in the next subsection.

### 3.2.1. The Continuity Equation in Logarithmic Form

A useful and stable method for solving this equation is to transform the equation into Logarithmic form. We start from (11) and first define $n_p(x)/n_{p0} = \exp[N(x)]$. Then we have

$$\frac{n_p'(x)}{n_p(x)} = N'(x) \tag{11}$$

$$\frac{n_p''(x)}{n_p(x)} - \left[\frac{n_p'(x)}{n_p(x)}\right]^2 = N''(x) \tag{12}$$

Now, by dividing both sides by $n_p(x)$, we get

$$-\left(1 - \frac{n_{p0}}{n_p(x)}\right) + L_n^2\left(\frac{E}{V_T}\frac{n_p'(x)}{n_p(x)} + \frac{n_p''(x)}{n_p(x)}\right) = 0 \tag{13}$$

where $L_n = \sqrt{\tau_n \mu_n V_T}$. Hence, we may rewrite (10) as the second-order nonlinear equation

$$\exp(-N) - 1 + L_n^2\left[\frac{E}{V_T}N' + (N'' + N'^2)\right] = 0$$
$$N(x=0) = V/V_T \tag{14}$$
$$N(x=W) = 0$$

We now define two parameters $M = N'$ and $f = E/V_T$. Using these parameters we obtain the nonlinear first-order system of equations

$$\begin{bmatrix} N \\ M \end{bmatrix}' = \begin{bmatrix} M \\ M^2 + fM + \dfrac{1-\exp(-N)}{L_n^2} \end{bmatrix} \tag{15}$$

which can be solved by the well-known Runge-Kutta method. The solutions in both exponential and logarithmic forms agree, but the logarithmic form is far more stable.

### 3.2.2. Approximate Solution to the Continuity Equation

Here, we suggest an analytical approximation to (9,10). First, we suppose that the electric field at each point located in the neutral region is a constant parameter. Then we solve the equation just if as the electric field would have zero derivatives. Once the expression is found, the position-dependent electric field is plugged in. Obviously, this method is sufficiently accurate for slowly varying electric fields. Thus, this approximation gives

$$p_n(x) - p_{n0} \approx p_{n0}\left[\exp(V/V_T) - 1\right]$$
$$\times \frac{\exp[x/L_{p2}(x)]\exp[X_n/L_{p1}(x)] - \exp[x/L_{p1}(x)]\exp[X_n/L_{p2}(x)]}{\exp[x_{n0}/L_{p2}(x)]\exp[X_n/L_{p1}(x)] - \exp[x_{n0}/L_{p1}(x)]\exp[X_n/L_{p2}(x)]} \tag{16}$$

where $L_{pi}(x) = 2\tau_p V_T\left[\tau_p E(x) - (-1)^i \sqrt{\tau_p^2 E^2(x) + 4\tau_p V_T/\mu_p}\right]^{-1}$.

Similarly, we obtain the injected holes distribution in the p-side theoretically as

$$n_p(x) - n_{p0} \approx n_{p0}\left[\exp(V/V_T) - 1\right]$$
$$\times \frac{\exp[-x/L_{n1}(x)]\exp[X_p/L_{n2}(x)] - \exp[-x/L_{n2}(x)]\exp[X_p/L_{n1}(x)]}{\exp[x_{p0}/L_{n2}(x)]\exp[X_p/L_{n1}(x)] - \exp[x_{p0}/L_{n1}(x)]\exp[X_p/L_{n2}(x)]} \tag{17}$$

where $L_{ni}(x) = 2\tau_n V_T\left[\tau_n E(x) - (-1)^i \sqrt{\tau_n^2 E^2(x) + 4\tau_n V_T/\mu_n}\right]^{-1}$.

In practice, these approximations have been verified against numerical solutions of the exponential and logarithmic nonlinear differential equations, yielding surprisingly good agreements, as discussed later below.

Now, we proceed to design of the step 2D p-n junction. The overall device length is taken to be about 370nm and the surface dopant concentrations are $N_D = 1\times 10^{12}$cm$^{-2}$ and $N_A = 5\times 10^{11}$cm$^{-2}$ as depicted in figure 2. All calculations are here done per unit-width

**Table 1.** Numerical values of hole and electron current densities.

| $J_{p0}$(pA/cm) | $J_{n0}$(pA/cm) |
|---|---|
| 0.332 | 0.310 |

of the device. Using (8), the depletion width is found to be 168.5nm.

Figure 3 illustrates the corresponding electric field and potential distributions. Based on this electric field distribution, we solve (9,10) both numerically and using the approximate formulae (16,17). Figure 4 shows how the exact numerical solution (dashed curves) and approximate theoretical formulae (solid curves) behave very closely. The results prove the acceptable accuracy of the formulas.

Now, we continue to derive the current densities. At $x = x_{n0}$, the hole current density is

$$J_p = q\mu_p(E - V_T \frac{d}{dx})p_n \bigg|_{x=x_{n0}} = J_{p0}\left[\exp(V/V_T) - 1\right]$$

$$J_{p0} = q\mu_p p_{n0} E(x_{n0}) - \frac{q\mu_p p_{n0} V_T}{A(x_{n0})} \left[ \frac{\exp[x_{n0}/L_{p1}(x_{n0})]\exp[X_n/L_{p2}(x_{n0})]}{L_{p1}(x_{n0})} \right.$$

$$\left. - \frac{\exp[x_{n0}/L_{p2}(x_{n0})]\exp[X_n/L_{p1}(x_{n0})]}{L_{p2}(x_{n0})} \right] \quad (18)$$

$$A(x_{n0}) = \exp\left[\frac{x_{n0}}{L_{p1}(x_{n0})}\right]\exp\left[\frac{X_n}{L_{p2}(x_{n0})}\right] - \exp\left[\frac{x_{n0}}{L_{p2}(x_{n0})}\right]\exp\left[\frac{X_n}{L_{p1}(x_{n0})}\right]$$

Similarly, we obtain the electron current in the p-side at $x = -x_{p0}$ as

$$J_n = q\mu_n(E + V_T \frac{d}{dx})n_p \bigg|_{x=-x_{p0}} = J_{n0}\left[\exp(V/V_T) - 1\right]$$

$$J_{n0} = q\mu_n n_{p0} E(-x_{p0}) - \frac{q\mu_n n_{p0} V_T}{A(x_{p0})} \left[ \frac{\exp[x_{p0}/L_{n1}(-x_{p0})]\exp[X_p/L_{n2}(-x_{p0})]}{L_{n1}(-x_{p0})} \right.$$

$$\left. - \frac{\exp[x_{p0}/L_{n2}(-x_{p0})]\exp[X_p/L_{n1}(-x_{p0})]}{L_{n2}(-x_{p0})} \right] \quad (19)$$

$$A(x_{p0}) = \exp\left[\frac{x_{p0}}{L_{n1}(-x_{p0})}\right]\exp\left[\frac{X_p}{L_{n2}(-x_{p0})}\right] - \exp\left[\frac{x_{p0}}{L_{n2}(-x_{p0})}\right]\exp\left[\frac{X_p}{L_{n1}(-x_{p0})}\right]$$

Table 1 shows the numerical values of the hole and electron current densities expected for this p-n junction. The total current density is given by the sum of the (18) and (19) as

$$J = J_p + J_n = b + a\left[\exp(\frac{V}{V_T}) - 1\right] = (J_{p0} + J_{n0})\left[\exp(\frac{V}{V_T}) - 1\right] \quad (20)$$

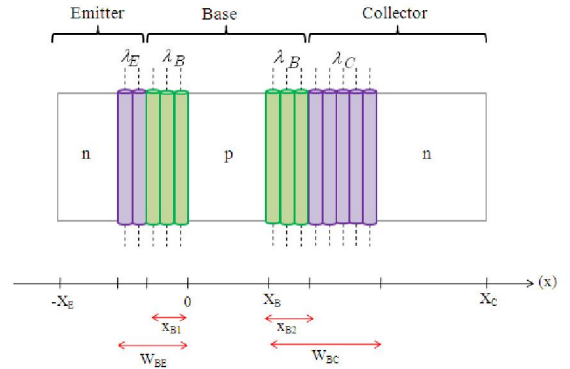

**Figure 5.** Electrostatic model for n-p-n transistor to approximate planar charge distribution using line charges.

### 4. The n-p-n Bipolar Transistor

In this section, we design and analyze an n-p-n 2D-BJT with abrupt charge doping profile and finite width. Figure 5 shows the electrostatic model to approximate planar charge distribution using line charges with infinitesimal width and charge density. Based on this charge model the unified equation for the total electric field over the surface of the device is given by

$$E(x) = -\lambda_E \ln\left|\frac{x + x_{B1}}{x + W_{BE}}\right| + \lambda_B \ln\left|\frac{x(x - (X_B + x_{B2}))}{(x + x_{B1})(x - X_B)}\right|$$

$$- \lambda_C \ln\left|\frac{x - (X_B + W_{BC})}{x - (X_B + x_{B2})}\right| \quad (21)$$

where $\lambda_\mu = N_\mu/2\pi\varepsilon, \mu = E, B, C$.

Figure 6 illustrates the electric field and potential distributions over the surface of our designed 2D-BJT with a total width of about 7.7µm. Here, the dopant concentrations are $N_E=1\times10^{12}$cm$^{-2}$, $N_B=1\times10^{11}$cm$^{-2}$ and $N_C=1\times10^{10}$cm$^{-2}$. The lengths of terminals are supposed to be 124nm, 464nm and 7.1µm for emitter, base and collector respectively. Based on (8), the depletion widths are found to be $x_{B1}$=242.5nm, $W_{BE}$=266.75nm, $x_{B2}$=211.58nm, $W_{BC}$=2.33µm as shown in Figure 6.

Now, we may proceed to obtain the basic parameters introduced for a planar 2D bipolar transistor.

#### 4.1. Current Gain

Figure 7 presents a schematic of an n-type 2D-BJT, connected in an external bias circuit. We here assume a common base configuration and normal (active) mode of operation to find the corresponding current gain. Calculation of current gain would obviously need a thorough analysis of carrier transport across the width of structure.

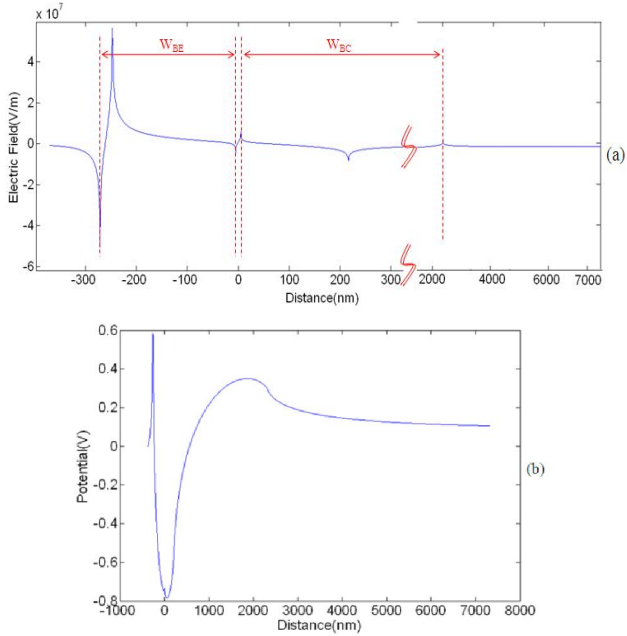

**Figure 6.** (a) Electric field distribution; (b) Electric potential distribution for the mentioned n-p-n transistor.

All the terminal currents are partitioned into hole and electron components, and then from the continuity equation we obtain the steady-state currents.

The transport in the neutral base region of the device is non-trivial and contains both diffusion and drift terms. Hence, the distribution of injected minority-carriers (electrons) in the base region is governed by the continuity conditions at the two edges, subject to the Shockley's boundary conditions

$$-\frac{1}{\tau_n}(n_p - n_{p0}) + \mu_n \left(E\frac{d}{dx} + V_T \frac{d^2}{dx^2}\right) n_p = 0$$
$$n_p(x=0) = n_{p0} \exp(V_{BE}/V_T)$$
$$n_p(x=X_B) = n_{p0} \exp(V_{BC}/V_T)$$
(22)

Figure 8 shows the electron distribution in the neutral base region based on numerical solution of this equation.

The electron surface current densities at the emitter edge $J_{nE}$ and the collector edge $J_{nC}$ are given by

$$J_{nE} = q\mu_n (E + V_T \frac{d}{dx}) n_p \Big|_{x=0}$$
$$= a_{nE}\left[\exp(V_{BE}/V_T) - 1\right] - b_{nE}\left[\exp(V_{BC}/V_T) - 1\right]$$
(23)

$$J_{nC} = q\mu_n (E + V_T \frac{d}{dx}) n_p \Big|_{x=X_B}$$
$$= a_{nC}\left[\exp(V_{BE}/V_T) - 1\right] - b_{nC}\left[\exp(V_{BC}/V_T) - 1\right]$$
(24)

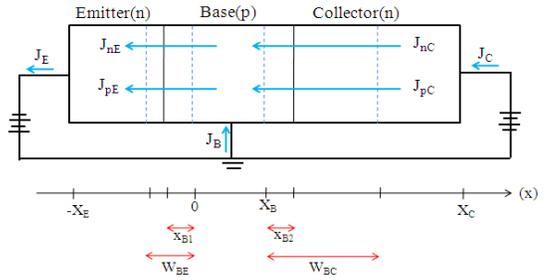

**Figure 7.** An n-p-n transistor under the normal forward operating conditions.

Similarly, in the pseudo-neutral emitter and collector regions, the equations for minority-carriers $p_n$ are

$$-\frac{1}{\tau_p}(p_{nE(C)} - p_{n0}) - \mu_p \left(E\frac{d}{dx} + V_T \frac{d^2}{dx^2}\right) p_{nE(C)} = 0$$
$$p_{nE(C)}(x = x_{0E(C)}) = p_{n0E(C)} \exp(V_{BE(BC)}/V_T)$$
$$p_{nE(C)}(x = x_{endE(C)}) = p_{n0E(C)}$$
(25)

with $x_{0E} = -W_{BE}$, $x_{0C} = X_B + W_{BC}$, $x_{endE} = -X_E$ and $x_{endC} = X_C$ as illustrated in figure 7 Numerical solutions of these equations are shown in figure 8.

The hole surface current densities at the emitter edge $J_{pE}$ and the collector edge $J_{pC}$ are given by

$$J_{pE(C)} = q\mu_{pE(C)} (p_{nE(C)} E - V_T \frac{d}{dx} p_{nE(C)}) \Big|_{x=x_{0E}(x_{0C})}$$
$$= a_{pE(C)} \left(\exp(V_{BE(BC)}/V_T) - 1\right)$$
(26)

Numerical parameter values for the constant parameters in the above sets of equations (23,24,25) are enlisted in Table 2.

Now, the terminal surface current densities can be summed over respective electron and hole currents as $J_E = J_{nE} + J_{pE}$, $J_C = J_{nC} + J_{pC}$, and $J_B = J_E - J_C$. Hence, the common-base and common-emitter current gains are found to be

$$\alpha = \frac{J_C}{J_E} = \frac{J_{nE}}{J_E} \frac{J_{nC}}{J_{nE}} \frac{J_C}{J_{nC}} = 0.993$$
(27)

$$\beta = \frac{\alpha}{1-\alpha} = 138$$
(28)

Figure 9 shows $J_C$ versus $V_{CE}$ for the common-emitter configuration under normal and inverted bias configuration. The saturation region is here defined as $V_{CE} < V_{CE,sat} = 0.1V$ and the current gain in inverted mode is found to be $\beta_r = 0.9$. In contrast, standard bulk BJTs start to enter the saturation at $V_{CE} < 0.2V$, and therefore 2D-BJTs may operate at smaller voltages comparing to their bulk counterparts.

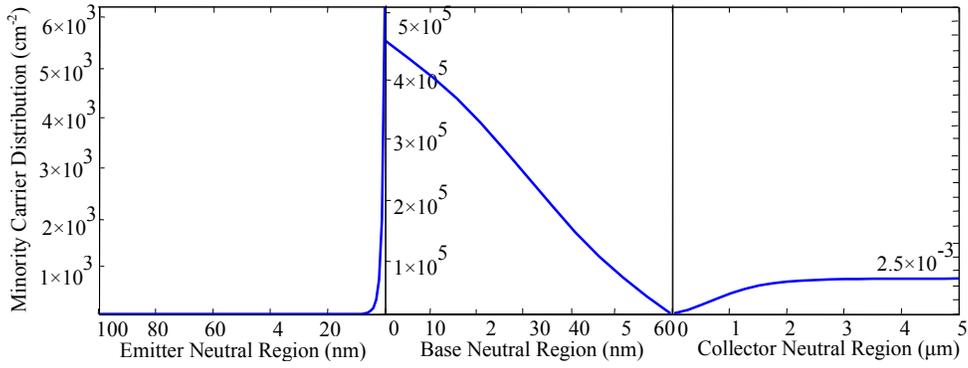

**Figure 8.** Minority carries distributions in neutral regions of the emitter, base and collector.

**Table 2.** Numerical values of constants for hole and electron current densities.

| $J_{nE}$(pA/cm) | | $J_{nC}$(pA/cm) | | $J_{pE}$(pA/cm) | $J_{pC}$(pA/cm) |
|---|---|---|---|---|---|
| $a_{nE}$ | $b_{nE}$ | $a_{nC}$ | $b_{nC}$ | $a_{pE}$ | $a_{pC}$ |
| 2.04 | 3.9×10$^{-2}$ | 2.03 | 3.91×10$^{-2}$ | 1.4×10$^{-2}$ | 9.16×10$^{-3}$ |

### 4.2. Small-Signal Model

The equivalent circuit for the behavior of this planar BJT is identical to that of a bulk BJT. Small-signal model parameters are here defined as transconductance $g_m = \partial i_C/\partial v_{BE}$, output resistance $r_o$, and input resistance $r_\pi = \partial i_B/\partial v_{BE} = \beta/g_m$. Based on the flatness of the $I_C$ curves, the Early voltage $V_A$ is obtained to be infinity, and hence the output resistance expressed as $r_o \equiv \left(\partial i_C/\partial v_{CE}\right)^{-1} = V_A/I_C$ was obtained as virtually unlimited. This is particularly important for obtaining very large voltage gains at low operating voltages and power consumptions using properly designed p-n-p 2D-BJTs as active loads, and also for designing near-ideal current-sources.

The diffusion capacitance represents the injected electron concentration versus distance in the neutral base given as

$$C_\pi = q\int \frac{\partial n(x,V)}{\partial V} dx \qquad (29)$$

Variation of $C_\pi$ as a function of applied voltage is shown in Figure 10. The general trend is that it attains a maximum value of roughly $C_{\pi,\max} = 32(nF/cm)$ at a bias of about $V_{BE} = 0.82(V)$, and starts to fall afterwards. This behavior is quite typical and also seen in bulk BJTs as well. Also, the uncompensated dopants in the transition regions of a p-n junction cause dipoles, causing the junction capacitance

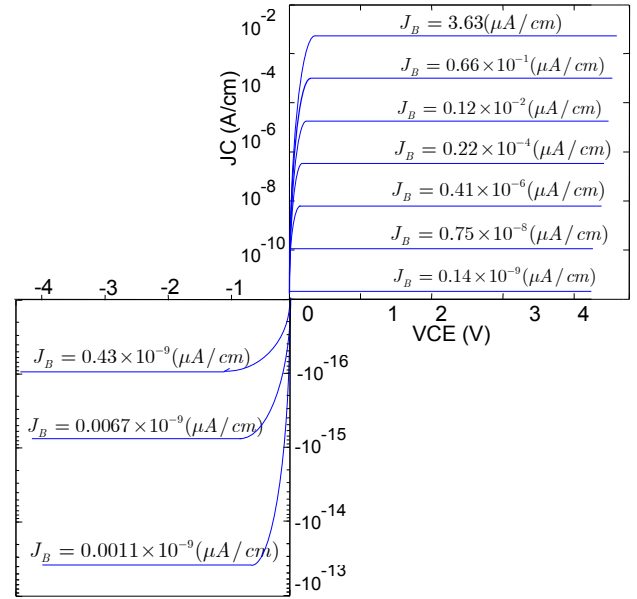

**Figure 9.** Output characteristics in common emitter configuration.

$$C_j = \left|dQ/dV\right| \qquad (30)$$

Figure 11 also depicts the junction capacitance of the reverse-biased B-C junction versus $V_{BC}$.

Now we assume that an AC signal is fed into the device by a current source $i_b$ connected to the base. The cutoff frequency $f_T$ is the frequency at which the AC current gain $\beta_{ac} = i_c/i_b$ falls to unity. Figure 11 shows the variation of $f_T$ as a function of collector surface current density $J_C$. The maximum value of this frequency is about 77GHz at a fairly surface current density of 0.38 mA/μm.

The typical consumed power is very small. A comparison to an existing design [23] of bulk BJTs may be made by taking an identical base width of 1μm. Then at the maximum operating frequency which corresponds to a transit time of $\tau_t = 1/(2\pi f_T) = 2.1 ps$, and taking a collector bias voltage of 5V, then operating power (per 1μm width) would be around 1.9mW. Hence, the estimated power-delay product [24] will be around 4fJ, far less than the typical record values which are of the order of tens of femto-Joules.

**Table 3.** Numerical values of Ebers-Moll parametres.

| $J_{ED}$ (pA/cm) | $J_{CD}$ (pA/cm) | $\alpha_N$ | $\alpha_I$ |
|---|---|---|---|
| 2.04 | 2.03 | 0.995 | 0.98 |

It has to be mentioned that this power-delay product has been derived when the 2D-BJT is such biased to allow the maximum operation speed. As it has been shown in Figure 12, much smaller power-delay products can be expected if lower bias current and therefore the threshold frequency is chosen. Actually, the power-delay product blows up at large bias currents, and for instance if one limits the cutoff frequency to 70GHz, the resulting power-delay product will be strictly limited to the record value of 0.01fJ/μm.

*4.3. Large-Signal Model*

The DC emitter and collector currents in active mode are well modeled by an approximation to the Ebers–Moll model. We have thus the relationships

$$J_E = J_{ED}\left[\exp(V_{BE}/V_T)-1\right] - \alpha_I J_{CD}\left[\exp(V_{BC}/V_T)-1\right] \quad (31)$$

$$J_C = \alpha_N J_{ED}\left[\exp(V_{BE}/V_T)-1\right] - J_{CD}\left[\exp(V_{BC}/V_T)-1\right] \quad (32)$$

Numerical parameter values for these set of equations are enlisted in Table 3.

## 5. Approximate Formulae for an n-p-n 2D-BJT

Here we will derive parameters of n-type 2D-BJTs based on the theory derived for p-n junction in section III. A dual analysis will be evidently applicable to p-type 2D-BJTs by changing the carrier types and voltage and current polarities.

For calculation of transistor parameters, the injected minority-carriers distributions are necessary, and here we present the approximation solutions following the equations calculated in the above sections.

Starting with the minority-carriers in the base, we have to solve (23). The approximate solution to this equation is

$$n_{pB}(x) - n_{p0B} \approx$$
$$\frac{\exp\left[-x/L_{B2}(x)\right] - \exp\left[-x/L_{B1}(x)\right]}{\exp\left[-X_B/L_{B2}(x)\right] - \exp\left[-X_B/L_{B1}(x)\right]} n_{p0}\left[\exp(V_{BC}/V_T)-1\right]$$
$$+ \frac{\exp\left[-x/L_{B1}(x)\right]\exp\left[-X_B/L_{B2}(x)\right] - \exp\left[-x/L_{B2}(x)\right]\exp\left[-X_B/L_{B1}(x)\right]}{\exp\left[-X_B/L_{B2}(x)\right] - \exp\left[-X_B/L_{B1}(x)\right]}$$
$$\times n_{p0}\left[\exp(V_{BE}/V_T)-1\right] \quad (33)$$

where $L_{nBi}(x) = 2\tau_n V_T \left[\tau_n E(x) - (-1)^i \sqrt{\tau_n^2 E^2(x) + 4\tau_n V_T/\mu_n}\right]^{-1}$.

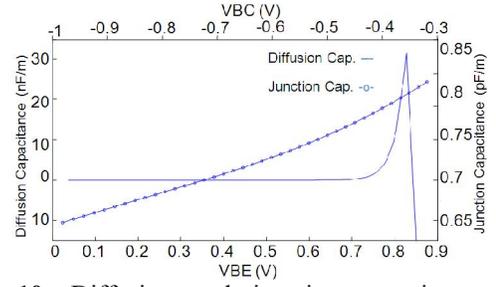

**Figure 10.** Diffusion and junction capacitance versus voltage.

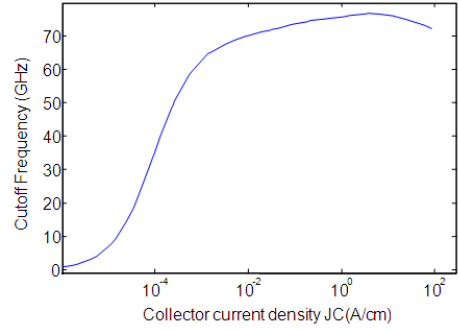

**Figure 11.** Cutoff frequency versus current density.

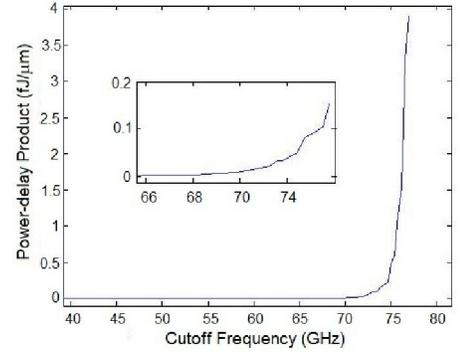

**Figure 12.** Power-delay versus cutoff frequency; the inset shows a magnification of the knee.

The electron surface current density of the emitter is similarly given by

$$J_{nE} = a_{nE}\left[\exp\left(V_{BE}/V_T\right)-1\right] + b_{nE}\left[\exp\left(V_{BC}/V_T\right)-1\right]$$

$$a_{nE} = q\mu_n n_{p0} E(0) - \frac{q\mu_n n_{p0} V_T}{\exp[-X_B/L_{B2}(0)] - \exp[-X_B/L_{B1}(0)]}$$

$$\times\left[\frac{1}{L_{B1}(0)}\exp[-X_B/L_{B2}(0)] - \frac{1}{L_{B2}(0)}\exp[-X_B/L_{B1}(0)]\right]$$

$$b_{nE} = \frac{q\mu_n n_{p0} V_T}{\exp[-X_B/L_{B2}(0)] - \exp[-X_B/L_{B1}(0)]}\left(\frac{1}{L_{B1}(0)} - \frac{1}{L_{B2}(0)}\right)$$

(34)

The electron surface current density of the collector will be

$$J_{nC} = a_{nC}\left[\exp\left(V_{BE}/V_T\right)-1\right] + b_{nC}\left[\exp\left(V_{BC}/V_T\right)-1\right]$$

$$a_{nC} = q\mu_n n_{p0} V_T \frac{\exp[-X_B/L_{B2}(X_B)]\exp[-X_B/L_{B1}(X_B)]}{\exp[-X_B/L_{B2}(X_B)] - \exp[-X_B/L_{B1}(X_B)]}$$

$$\times\left[\frac{1}{L_{B2}(X_B)} - \frac{1}{L_{B1}(X_B)}\right]$$

$$b_{nC} = q\mu_n n_{p0} E(X_B) + \frac{q\mu_n n_{p0} V_T}{\exp[-X_B/L_{B2}(X_B)] - \exp[-X_B/L_{B1}(X_B)]}$$

$$\times\left[\frac{1}{L_{B1}(X_B)}\exp[-X_B/L_{B2}(X_B)] - \frac{1}{L_{B2}(X_B)}\exp[-X_B/L_{B1}(X_B)]\right]$$

(35)

The hole distributions in the emitter and collector are obtainable using (25). The solutions are like those obtained for a p-n junction, given by

$$p_{nE(C)}(x) - p_{n0E(C)} \approx \frac{p_{n0E(C)}}{A(x)}\left[\exp(V_{BE(BC)}/V_T) - 1\right]$$

$$\times\left[\exp[x/L_{pE(C)2}(x)]\exp[x_{endE(C)}/L_{pE(C)1}(x)] - \exp[x/L_{pE(C)1}(x)]\exp[x_{endE(C)}/L_{pE(C)2}(x)]\right]$$

$$A(x) = \exp\left[\frac{x_{0E(C)}}{L_{pE(C)2}(x)}\right]\exp\left[\frac{x_{endE(C)}}{L_{pE(C)1}(x)}\right] - \exp\left[\frac{x_{0E(C)}}{L_{pE(C)1}(x)}\right]\exp\left[\frac{x_{endE(C)}}{L_{pE(C)2}(x)}\right]$$

(36)

with $L_{pE(C)i}(x) = 2\tau_p V_T\left[\tau_p E(x) - (-1)^i\sqrt{\tau_p^2 E^2(x) + 4\tau_p V_T/\mu_p}\right]^{-1}$.

Also, we have $x_{0E} = -W_{BE}$, $x_{0C} = X_B + W_{BC}$, $x_{endE} = -X_E$ and $x_{endC} = X_C$, as illustrated in Figure 7. The hole surface current densities in the emitter and collector are obtained as

$$J_{pE(C)} = a_{pE(C)}\left[\exp(V_{BE(BC)}/V_T) - 1\right]$$

$$a_{pE(C)} = q\mu_p p_{n0E(C)} E - \frac{q\mu_p n_{p0E(C)} V_T}{A(x_{0E(C)})} B(x_{0E(C)})$$

$$B(x_{0E(C)}) = \frac{1}{L_{E(C)1}(x_{0E(C)})}\exp[x_{0E(C)}/L_{E(C)1}(x_{0E(C)})]\exp[x_{endE(C)}/L_{E2}(x_{0E(C)})]$$

$$- \frac{1}{L_{E(C)2}(x_{0E(C)})}\exp[x_{0E(C)}/L_{E(C)2}(x_{0E(C)})]\exp[x_{endE(C)}/L_{E(C)1}(x_{0E(C)})]$$

(37)

## 5. Conclusions

We derived necessary formulae in studying 2D step p-n junctions and suggested an approximate analytical solution for continuity equations, which exhibited good agreement with exact numerical solutions. Then, we proceeded to develop the theory of 2D BJTs, and used the one-sided hydrogenated Graphene, or Graphone, as the example 2D material. Distributions of electric field and potential were calculated, and the small signal hybrid-$\pi$ model of the transistor was obtained. The results showed that this 2D-BJT would operate somewhat similar to bulk BJTs in DC and AC circuits, however, with very different performance parameters. The maximum frequency and current gain were found to be respectively as 77GHz and 138, at a maximum power-delay product of 4fJ/μm. This is while much smaller power-delay products would be possible at the expense of less threshold frequency.

This preliminary analysis demonstrates the usefulness of such 2D-BJTs, in several aspects. As we have shown, a 2D-BJT would allow appreciable current gain, large operating frequencies, yet very small power-delay products which may be ultimately attributed to the associated ultrathin 2D materials. Furthermore, when used as active loads in a properly designed circuit, 2D-BJTs may be able to outperform their CMOS counterparts with regard to the large voltage gains at exceedingly small power consumption.